\documentclass{article}
\usepackage{graphicx}
\usepackage{amsfonts}
\usepackage{amsmath}
\usepackage{amssymb}
\usepackage{hyperref}
\usepackage{indentfirst}
\usepackage[usenames,dvipsnames]{pstricks}
\usepackage{epsfig}
\usepackage{pst-grad} 
\usepackage{pst-plot} 

\title{Ground-state configurations in ferromagnetic nanotori}
\author{V. L. Carvalho-Santos $^{1,2,\,}$\footnote{E-mail address: vagson.santos@ufv.br}  ,
W. A. Moura-Melo$^{1,\,}$\footnote{E-mail address: winder@ufv.br} ,
A. R. Pereira$^{1,3,\,}$\footnote{E-mail adress: apereira@ufv.br, pereira@pks.mpg.de}
\\
\it\small $^{1}$ Departamento de F\'isica, Universidade Federal de
Vi\c cosa\\\it\small 36570-000, Vi\c cosa, Minas Gerais, Brazil\\
\it\small $^{2}$ Instituto Federal de Educa\c c\~ao, Ci\^encia e Tecnologia Baiano - Campus Senhor do Bonfim \\\it\small 48970-000, Senhor do Bonfim, Bahia, Brazil\\
\it\small $^{3}$ Max Planck Institute for the Physics of Complex Systems, 01187 Dresden, Germany}
\date{}

\begin{document}
\maketitle
\begin{abstract}Magnetization ground states are studied in toroidal nanomagnets. The energetics associated to the ferromagnetic, vortex and onion-like configurations are explicitly computed. The analysis reveals that the vortex appears to be the most prominent of such states, minimizing total energy in every torus with internal radius $r\gtrsim10\,{\rm nm}$ (for Permalloy). For $r\lesssim10\,{\rm nm}$ the vortex remains the most favorable pattern whenever $R/\ell_{ex}\gtrsim1.5$ ($R$ is the torus external radius and $\ell_{ex}$ is the exchange length), being substituted by the ferromagnetic state whenever $R/\ell_{ex}\lesssim1.5$.\end{abstract}

\section{Introduction and Motivation}
Nanomagnetism has become an important area for both basic and applied research in the last years. Such an importance is greatly due to recent advances in the synthesis and characterization of magnetic samples of very small sizes, comprising micro and sub-micrometer scales. Nanomagnets of several shapes and sizes have appeared and have been intensively investigated, namely for potential applications in logic memory, data storage and highly sensitive sensors \cite{nanoapplications}. Such an interest has been renewed and reinforced by virtue of a very recent proposal of using biofunctionalized magnetic-vortex microdiscs for triggering cancer therapy \cite{Cancer-therapy}. Often, such applications are based upon new magneto-electronic mechanisms which preclude on demand control of the magnetization stable configurations, like single-domains, domain-walls, vortices, etc. Among other, cylindrical and rectangular-type shape samples appear to be the most studied ones, once their relatively simple geometries make easier the fabrication and eventually the comparison between experimental and theoretical findings. For instance, in a thin enough sub-micron sized disc of soft material, like Co or Permalloy (FeNi), the remanent (at zero applied field) ground-state is observed to be a single-vortex of rotating dipoles with a central core where they revolve against the disc plane for regularizing exchange cost \cite{nanodisc-vortex}. As the disc thickness is raised the vortex is progressively distorted being eventually substituted by a single-domain configuration \cite{single-domain}. In turn, thin rectangular-like samples generally present Landau patterns as their remanent ground-states \cite{Landau-states}.\\

It is widely recognized that the total energy of a magnet with an arbitrary magnetization pattern is not easy to be computed, even assuming simple geometries and negligible anisotropies. With these simplifications, exchange and magnetostatic contributions remain to be evaluated \cite{Aharoni_Schaffer-books,Getzlaf-Skomski-books}. The latter one is generally much more complicated once it comes from long-range dipole-dipole interactions. For special configurations, like vortices and some aligned states, and/or simple geometries, like those curvatureless, say, planar-type and cylinder, its evaluation is greatly simplified and may be sometimes exactly computed. For more general situations, the available tools for doing that often relies on approximate methods, e.g., assuming the nanomagnet as a continuum sample (analytical treatments) and/or performing numerical evaluation of the energy \cite{Aharoni_Schaffer-books}. For the specific cases of the square-type and cylindrical shapes, considerable amount of information are available  with relatively good agreement between experimental and theoretical studies \cite{review-exp-th-disc-square}. For other geometries, a detailed knowledge is still lacking, although some works have dealt with nanosized magnetic spheres \cite{nanospheres}, wires \cite{nanowires}, tubes \cite{nanotubes} and rings \cite{nanorings,Bellegia-etal-JMMM301-131-2006,Zhu-etal-PRL96-027205-2006}.\\

Whenever designing novel magneto-electronic mechanisms based upon vortex-like states, namely those for logic memory and/or data storage, an issue of prime importance concerns miniaturization of each nanomagnet and consequently of the whole array comprising a large number of them. Since current transistor-based memory elements have reached below $100\, {\rm nm}$ per element, the departure to nanomagnetic vortex-based mechanisms should offer real potentiality of supporting such stable states at much smaller scales, say, at least around $20-30\, {\rm nm}$, or less. There lies a trouble with soft material nanodiscs, where vortex stability is lost if the disc radius gets lower than $\sim 45\, {\rm nm}$ (value for Permalloy disks with thickness $10\, {\rm nm}$) \cite{Cowburn-PRL83-1042-1999}. To partially remedy the situation, one may introduce a (centralized) hole into the disc, obtaining a nanoring, where a vortex-state may be stabilized at smaller radius ($\sim 20\, {\rm nm}$, for internal radius and thickness about $10\, {\rm nm}$) \cite{Bellegia-etal-JMMM301-131-2006}. Physically, the presence of the hole becomes the core formation no longer necessary, diminishing exchange cost considerably and giving extra stability to vortex configuration even at smaller scales, see Refs. \cite{Rahm,nossospapers}. However, disappearing with the core, we immediately loose the vortex polarization, remaining only its chirality (sense of dipole rotating, clock or counter-clockwise) as bit storage media. Additionally, dipole dynamics at the edge borders may still jeopardize vortex stability due to large fluctuations of surface magnetic charges, $\sigma_{mag}=\vec{M}\cdot \hat{n}$, once at these places the normal vector $\hat{n}$ is not well-defined (in other words, curvature blows up at disc and hole edge borders). Such fluctuations may be somewhat bypassed and their effects minimized if the centralized hole is large so that the nanoring width is sufficiently small, less than a domain-wall size, say, around a couple of exchange length ($\ell_{\rm ex} \sim 5.3\, {\rm nm}$, for Permalloy). Doing that we also ensure that no vortex core can be formed in the ring \cite{Zhu-etal-PRL96-027205-2006}.\\

Here, we would like to consider a number of magnetization patterns lying in the toroidal manifold. Specifically, we compute the net energy (exchange + magnetostatic) of ferromagnetic (single-domain, SD), vortex and onion-like states. Our main results may be divided according nanotori internal radius, $r$: for small values, $r\lesssim 10\,{\rm nm}$, SD appears to be the ground-state if $R/\ell_{ex}\lesssim 1.5$, being substituted by a vortex configuration whenever $R/\ell_{ex}\gtrsim 1.5$. Thus, even in a very small torus, say,  with $r=1\, {\rm nm}$, we may have a stable vortex if $R\gtrsim 7.5\, {\rm nm}$. On the other hand, if $r\gtrsim 10\, {\rm nm}$ only the vortex pattern appears to minimize the net energy, despite the aspect ratio. Such a vortex stability, even at very small scales, should be attributed to the toroidal geometry, where curvature varies smoothly, preventing large fluctuations of surface charges at the edges, as may occur in nanorings. A direct consequence concerns the possibility of great improvement in miniaturization using stable vortices lying in tori as media for binary storage and logic. Note that such vortices are coreless, leaving only their chirality at our disposal. However, the vortex core absence is an important property when we consider applications where non-interacting magnetic elements concern. For example, the reduced dipole-dipole interaction may, in this situation, increases the vortex-state stability for small intertori separation distances, whenever compared to its interdiscs counterpart. Such a feature may be important for fabricating arrays with higher density, using nanotori as the basic elements. For applications in cancer therapy, the vanishing magnetization in remanence is significant because this reduces the long-range magnetostatic forces responsible for particle agglomeration.\\

To our best knowledge, there is so far no work reporting the fabrication of magnetic nanotori. The best we have found in the literature concerns some possibilities of producing such structures using their carbon-made analogues as a sort of substrate where a suitable magnetic material could be formed either by capping the surface with a thin layer or by injecting the material inside to form a nanotorus encapsulated by the layer \cite{torus-fabrication-1a}. By means of this latter technique, it has been reported in Ref. \cite{torus-fabrication-1b} Permalloy-made nanowires with (internal) radius around $10\, {\rm nm}$  (other possibilities for nanowires fabrication may be found in Refs. \cite{torus-fabrication-2}). Once experimental techniques in nanomagnetism have developed rapidly it is expected that, among other, magnetic nanotori may appear in the near future.\\

At some extent, this work represents a further step from that presented in Ref. \cite{Vagson-PRB2008}, where we have studied solitons and vortices in the toroidal geometry considering only the exchange energy. Here, for dealing with nanostructures their magnetostatic contribution must be also taken into account. We outline our article as follows: in Section 2, we describe the theoretical model, the torus geometry and the magnetization states under consideration. Section 3 is devoted to the analytical results concerning the energy of these magnetic states. In Section 4, we evaluate numerically the previous findings, followed by a suitable discussion about them. Finally, we conclude our work and present a number of prospects for future investigation.\\

\section{The model and the toroidal magnetization states}
In our approach, a magnetic nanotorus is assumed to be a continuum material where the magnetization vector has the same magnitude throughout it, but can vary in direction from point to point. Its energy comes from short-range exchange ($E_{\rm ex}$) plus long-range dipole-dipole (magnetostatic, $E_{\rm mag}$) interactions, once we assume vanishing anisotropies, i.e., we are explicitly dealing with isotropic materials, like Permalloy (FeNi) and Co. Therefore, the ground-state will be the configuration which minimizes $E_{\rm net} = E_{\rm ex} + E_{\rm mag}$, which generally depends upon the torus geometry. Explicitly, we have that the exchange term is given by the continuum version of the Heisenberg model over the whole volume \cite{Vagson-PRB2008}:
\begin{equation}\label{Eex}
E_{\text{ex}}=A\int_{V}\left[\left(\vec{\nabla}m_{x}\right)^{2}+\left(\vec{\nabla}m_{y}\right)^{2}+\left(\vec{\nabla}m_{z}\right)^{2}\right]dV,
\end{equation}  
where $A$ accounts for the interaction strength among nearest-neighbor spins (dipoles). Once we are considering a ferromagnetic sample, $A>0$, so that $E_{\rm ex}\ge 0$; $\vec{m}(\vec{r})=(m_x,m_y,m_z)=\vec{M}/M_s$ is the normalized magnetization and $M_s$ is its satured value. For Permalloy, one has $A\approx 1.3\times10^{-11}\, {\rm J/m}$ and $M_s \approx 8.6\times 10^{5}\, {\rm A/m}$, so that, $\ell_{\rm ex}=\sqrt{ 2A/\mu_0 M^2_s}\approx5.3\, {\rm nm}$. Such a length accounts for the relative importance of each contribution, in such a way that exchange dominates at small distances, $d\lesssim \ell_{\rm ex}$, whereas magnetostatic takes over at larger ones, $d\gtrsim \ell_{\rm ex}$. This latter energy comes from the interaction experienced by a given dipole due to the remaining ones throughout the magnet. With the additional assumption that no electric current is present, $\vec{J}\equiv 0$, we formally obtain:
\begin{equation}\label{Emag}
E_{\text{mag}}=\frac{\mu_{0}M_{S}}{2}\int_{V}\vec{m}(\vec{r})\cdot \vec{\nabla}\Phi(\vec{r})dV\geqslant0,
\end{equation}
with the magnetotatic potential given by:
\begin{equation}\label{potencialmag}
\Phi(\vec{r})=-\frac{M_{S}}{4\pi}\int_{V'}\frac{\rho_m(\vec{r}')}
{|\vec{r}-\vec{r}'|}dV'+\frac{M_{S}}{4\pi}\int_{S'}\frac{\sigma_m(\vec{r}')}
{|\vec{r}-\vec{r}'|}dS'\,,
\end{equation}
where $\rho_m(\vec{r}')=\vec{\nabla'}\cdot\vec{m}(\vec{r}')$ and $\sigma_m(\vec{r}')=\hat{n'}\cdot\vec{m}(\vec{r}')$ are the volumetric and superficial densities of effective magnetic charges, respectively.\\

Since our interest concerns ferromagnetic tori, it is instructive to give a survey of its shape with the main features. Topologically, an ordinary torus is the simplest two-dimensional closed manifold having genus-1, a single hole, like depicted in Fig. \ref{torus1}. Depending on the relative sizes of its {\em internal}, $r$, and external, $R$, radii, three distinct standard types are obtained: ring, horn, or self-intersecting spindle torus, according $R>r$, $R=r$, or $R<r$, respectively (see, for instance, Ref.\cite{Vagson-PRB2008} and related references therein). The commonest one is the ring torus, resembling a {\em donut}, which will be the type studied here (the other ones, although mathematically interesting, seem to be rare in Nature). Geometrically, a torus is a surface whose curvature smoothly varies along the polar angle, $\theta$, from negative to positive values. For describing its geometry some coordinate systems are frequently used, among them, the spherical-type relates the Cartesian to the toroidal angles, as below:

\begin{figure}
\includegraphics[scale=0.56]{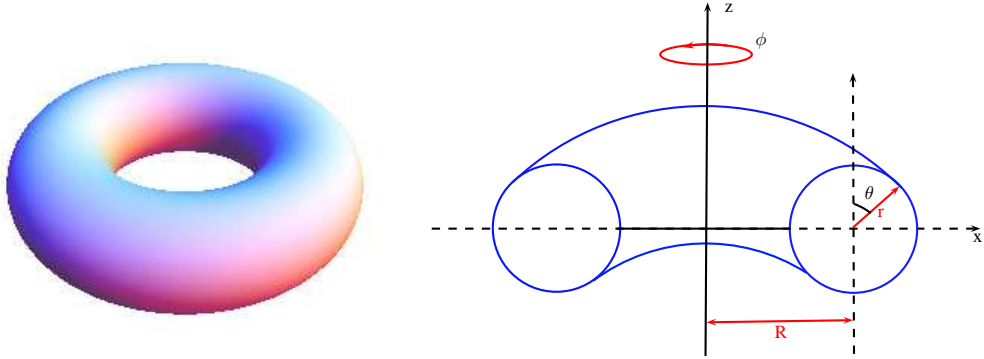}\hspace{-6em}\vspace{-2em}
\scalebox{0.7} {
\begin{pspicture}(0,-0.9238675)(12.83569,7.2838674)
\definecolor{color14}{rgb}{0.03137254901960784,0.0784313725490196,0.984313725490196}
\definecolor{color17}{rgb}{0.08627450980392157,0.10588235294117647,0.9882352941176471}
\definecolor{color19}{rgb}{0.0,0.12941176470588237,1.0}
\definecolor{color23}{rgb}{0.06666666666666667,0.09411764705882353,0.9372549019607843}
\definecolor{color28}{rgb}{0.0196078431372549,0.0,0.0}
\definecolor{color29}{rgb}{0.996078431372549,0.023529411764705882,0.023529411764705882}
\pscircle[linewidth=0.04,linecolor=color14,dimen=outer](4.6813145,2.9738674){1.23}
\psline[linewidth=0.04cm,linestyle=dashed,dash=0.16cm 0.16cm,arrowsize=0.05291667cm 2.0,arrowlength=1.4,arrowinset=0.4]{->}(2.3113146,2.9638674)(12.731315,2.9638674)
\pscircle[linewidth=0.04,linecolor=color17,dimen=outer](10.321315,2.9538674){1.23}
\rput{-44.622433}(2.786085,5.022816){\psarc[linewidth=0.04,linecolor=color19](7.513075,-0.8832874){3.5657797}{102.12629}{171.40565}}
\rput{-44.622433}(2.1580956,5.259048){\psarc[linewidth=0.04,linecolor=color23](7.486916,0.0){5.294164}{89.58384}{178.08524}}
\psline[linewidth=0.04cm,linecolor=red,arrowsize=0.05291667cm 2.0,arrowlength=1.4,arrowinset=0.4]{<->}(7.531315,1.1838675)(10.311315,1.2238675)
\psline[linewidth=0.04cm,linecolor=color28,linestyle=dashed,dash=0.16cm 0.16cm,arrowsize=0.05291667cm 2.0,arrowlength=1.4,arrowinset=0.4]{->}(10.331315,0.5838675)(10.311315,5.9238677)
\psline[linewidth=0.04cm,linecolor=color29,arrowsize=0.05291667cm 2.0,arrowlength=1.4,arrowinset=0.4]{->}(10.311315,2.9838674)(11.191315,3.7638674)
\usefont{T1}{ptm}{m}{n}
\rput(8.946628,0.9988675){\color{red}R}
\usefont{T1}{ptm}{m}{n}
\rput(10.843971,3.2588675){\color{red}r}
\usefont{T1}{ptm}{m}{n}
\rput(12.68569,2.7988675){x}
\usefont{T1}{ptm}{m}{n}
\rput(7.952877,7.1588674){z}
\rput{-30.894043}(0.19773728,5.4434905){\psarc[linewidth=0.04](9.948522,2.3639522){1.1368291}{82.7366}{101.401985}}
\psellipse[linewidth=0.04,linecolor=red,dimen=outer](7.5213146,6.2738676)(0.89,0.21)
\rput{-40.471733}(-0.9719867,5.941778){\psarc[linewidth=0.04,linecolor=red,arrowsize=0.05291667cm 2.1,arrowlength=1.6,arrowinset=0.4]{->}(7.5730753,4.289233){2.1804705}{120.97654}{143.51755}}
\psline[linewidth=0.04cm,arrowsize=0.05291667cm 2.0,arrowlength=1.4,arrowinset=0.4]{->}(7.511315,0.54386747)(7.551315,7.2638674)
\usefont{T1}{ptm}{m}{n}
\rput(8.565064,6.5388675){$\phi$}
\usefont{T1}{ptm}{m}{n}
\rput(10.625065,3.6588676){$\theta$}
\psline[linewidth=0.04cm](5.8313146,2.9638674)(9.131314,2.9638674)
\end{pspicture} 
}\caption{[Color online] Commonest kind of torus, the ring torus, and an useful coordinate system over it, $(\theta,\phi)\in [0,2\pi]$. Parameters $r$ and $R$ are the internal and external torus radii, respectively, whereas $R/r$ is the aspect ratio.}
\label{torus1}
\end{figure}
\begin{equation} \label{coord1}
x=(R+r\sin\theta)\cos\phi,\hspace{0.3cm}y=(R+r\sin\theta)\sin\phi\hspace{0.3cm}
\text{and} \hspace{1cm} z=r\cos\theta\,,
\end{equation}
with $(\theta,\phi)\in [0,2\pi]$ being the polar and azimuthal angles, respectively (see Fig. \ref{torus1}). Another useful coordinate system, along with a number of additional relations used for evaluating some quantities obtained in Section 3, may be found in Appendix {\bf A}.\\

Let us now discuss the main aspects concerning the relevant magnetization states whose energetics will be evaluated in Section 3. First, let us recall that the ferromagnetic state is reached when all the dipoles align along the same direction, leading to magnetization saturation. By virtue of the parallelism between neighbor dipoles, their exchange energy vanishes identically. In addition, no volumetric magnetic charges take place for this pattern, $\rho_m=\nabla\cdot\vec{m}\equiv0$, so that, only surface charges, $\sigma_m=\hat{n}\cdot\vec{m}\neq0$, contributes to its magnetostatic energy. Therefore, the results concerning a toroidal surface may be straightforward applied to a volumetric torus. In section 3, we shall compute the total energy for such a pattern lying along $z$ and $xy$-plane, given respectively by, $\vec{m}_{{\cal F}_z}=\hat{z}$ and $\vec{m}_{{\cal F}_{xy}}=\hat{x}$ (any other direction along $xy$ is energetically equivalent). As a matter of fact, we shall see that in any torus the $\vec{m}_{{\cal{F}}_{xy}}$-state is energetically favorable than its $z$ counterpart.\\

Another relevant state to be considered is that which minimizes magnetostatic term, which is achieved by vanishing both $\rho_m$ and $\sigma_m$. This corresponds to a curly pattern around the torus genus such that its spatial distribution ensures $\nabla\cdot\vec{m}\equiv0$, besides being perpendicular to the surface at every point, so that $\hat{n}\cdot\vec{m}\equiv0$. Such a vortex-state may be described by $\vec{m}_v=\pm \hat{\phi}$, where $\hat{\phi}$ is the unit vector along the azimuthal angle, whereas $\pm$ is associated to the vortex chirality, counter- or clockwise, respectively. Furthermore, it is easy to obtain that $\nabla(m_x + m_y)$ leading to a non-zero exchange contribution, given by eq. (\ref{Eex}), to be evaluated in Section 3.\\

\begin{figure}
{\includegraphics[scale=0.3]{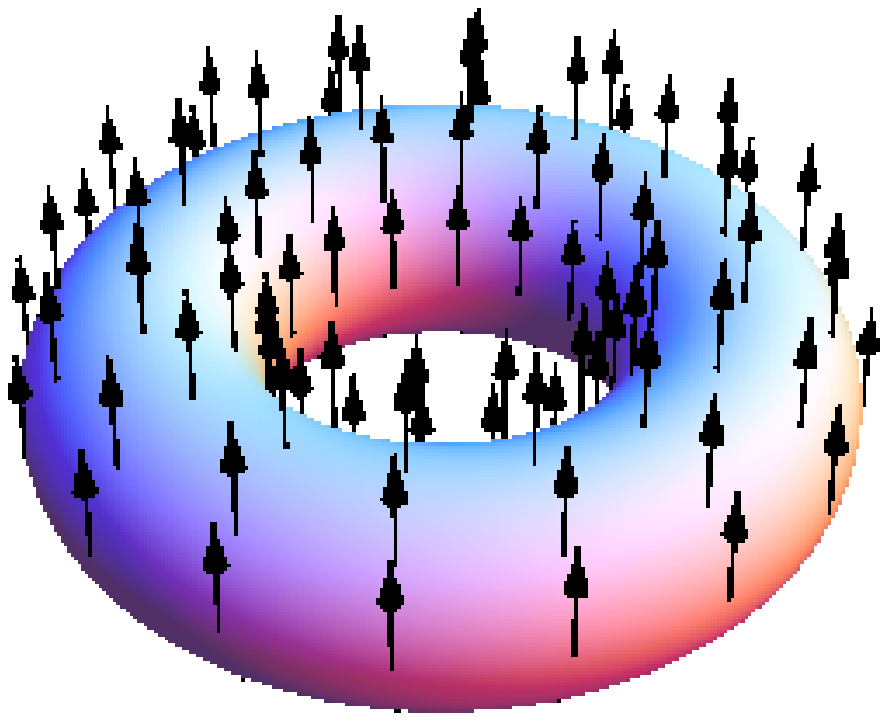}\includegraphics[scale=0.3]{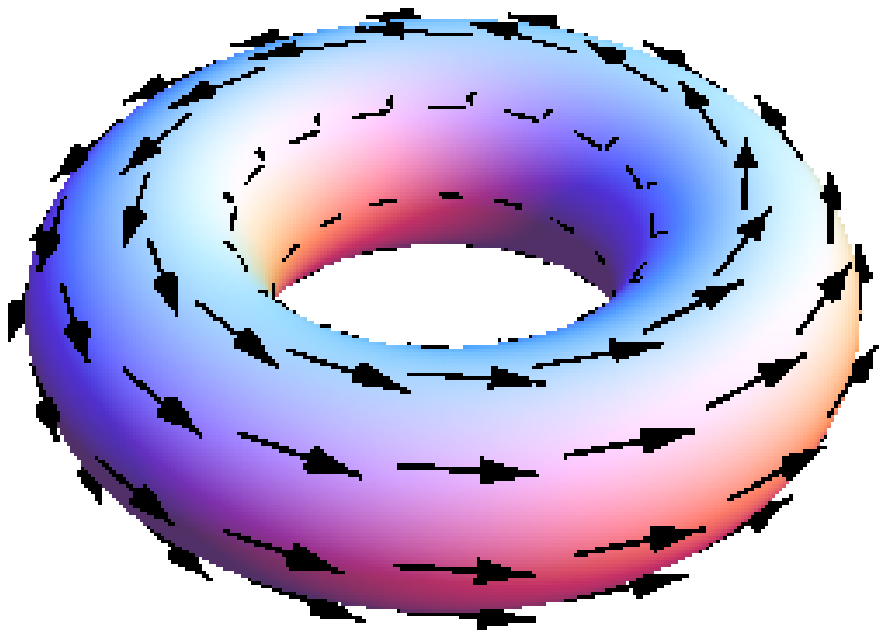}
\includegraphics[scale=0.3]{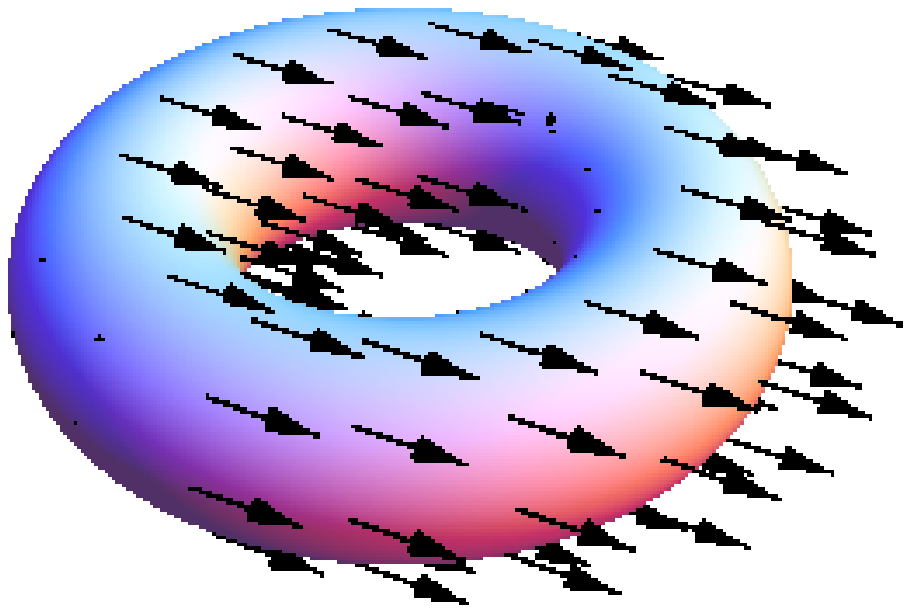}\includegraphics[scale=0.3]{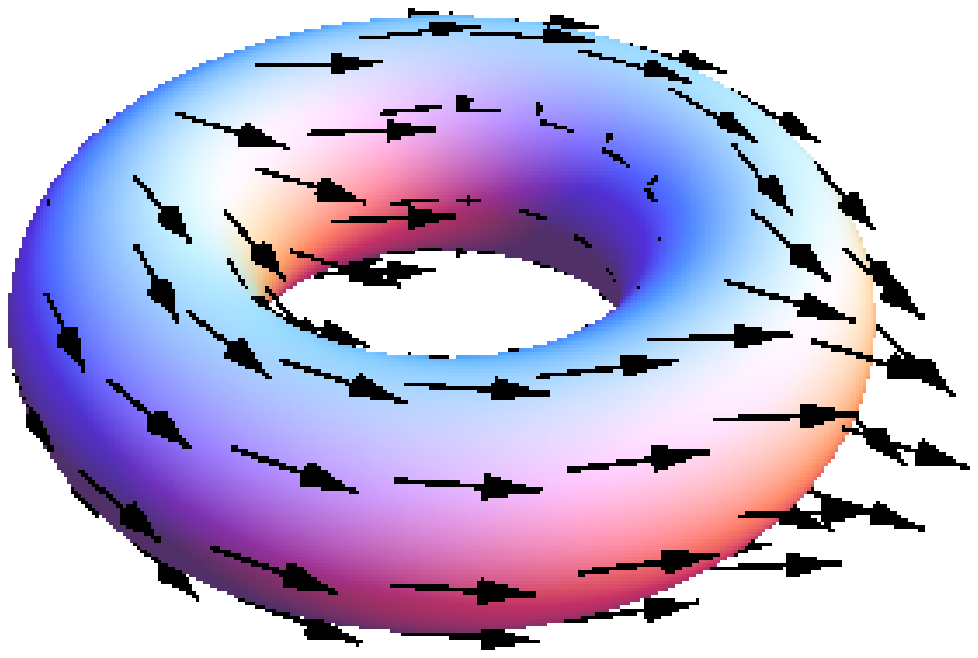}}
\caption{[Color online] Some configurations considered in this work. From left to right, one can see the single-domain (along $z$), the vortex,  ferromagnetic and onion (with $q=7$; see text for details) patterns, respectively (the last three ones with the dipoles lying along $xy$ plane).}
\label{Magnetic-states}
\end{figure}

Additionally, we also consider a more complex structure, known as an onion-like state. It may be  described by a magnetization depending on the azimuthal angle, $\vec{m}_{\cal O}=m_r(\phi)\hat{R} + m_\phi(\phi)\hat{\phi}$, where $\hat{R}=\hat{x}\cos\phi+\hat{y}\sin\phi$ and \cite{Landeros-JAP100-2006}:
\begin{equation}\label{onionstate}
\Big(m_{r},\,m_\phi\Big)=\left\{\begin{array}{cc}\Big(f(\phi),-\sqrt{1-f^2(\phi)}\Big), & 0<\phi<\pi/2\\\Big(-f(\pi-\phi),-\sqrt{1-f^2(\pi-\phi)}\Big), & \pi/2<\phi<\pi\\\Big(-f(\phi-\pi),\sqrt{1-f^2(\phi-\pi)}\Big), & \pi<\phi<3\pi/2\\\Big(f(2\pi-\phi),\sqrt{1-f^2(2\pi -\phi)}\Big), & 3\pi/2<\phi<2\pi\\\end{array}\right.
\end{equation}
with $f(q,\phi)\equiv f(\phi)= \cos^q(\phi) \, \in [-1,1]$. The real parameter $q\ge1$ accounts for the relative deviation of the pattern compared to the ferromagnetic state, $\vec{m}_{{\cal F}_{xy}}=\hat{x}$, formally recovered as long as $q\to1$ (Fig. \ref{Magnetic-states} shows a number of such configurations). Thus, as higher is the $q$ value more is the respective onion pattern distorted compared to the ferromagnetic state. These are the relevant configurations usually studied as possible ground-states in other nanomagnets, like discs, rings and spheres. Our intention here is to proceed with a similar plan in order to determine which of such states minimize the net energy for a given ring torus, whose specific geometry is dictated by its aspect ratio, $R/r$.\\

\section{The energetics of the configurations: analytical results}

\subsection{Exchange term}

As already said, ferromagnetic pattern has vanishing exchange cost, once magnetization does not vary throughout the sample, so that $\nabla\vec{m}\equiv0$.\\

Vortex exchange energy may be analytically computed within our approach based upon the continuum version of the Heisenberg model, eq. (\ref{Eex}), giving:
\begin{equation}\label{Eexvortex}
E_{\text ex}^{\mathcal{V}}=2\pi^{2}A\left[\sqrt{b^2 + t^{2}}-|b|\right],
\end{equation}
where $b^2=R^2-r^2>0$ and $|t|=\sqrt{r^{2}-c^{2}}\in(0,r]$ and $c$ is the hole radius of the hollow torus ($c=0$ represents the bulk torus). Expression above may be also obtained from its two-dimensional counterpart,  ${\cal E}_{\text ex}^{\mathcal{V}}=4\pi^{2}J r/\sqrt{R^2-r^2}$, by integrating over $r$ from $c$ to $r$, see Ref. \cite{Vagson-PRB2008} ($A=kJ/a$, where $a$ is the lattice spacing parameter, $J$ is the 2D coupling between the neighbour spins, the stiffness constant with energy dimension, and $k$ a real parameter depending upon the lattice structure, e.g., $k=1$ for simple cubic and $k=2$ for body-centered cubic crystals).\\

\begin{figure}
\begin{center}
\includegraphics[scale=0.4]{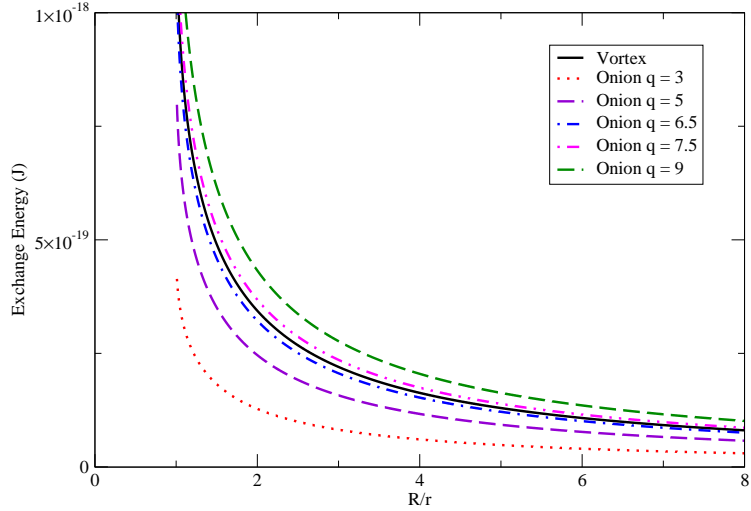}\caption{[Color online] Exchange energy as the function of the torus aspect ratio, $R/r$, for a number of possible configurations. Namely, note that whenever $q>7$ the onion-type is more energetic than the vortex state.}\label{Vortex-onion-Exch}
\end{center}
\end{figure}

Analogously, the onion-state exchange cost may be evaluated to give:
\begin{equation}\label{Eexonion}
E_{\text ex}^{\mathcal{O}}= E_{\text ex}^{\mathcal{V}}\Big( \mathcal{I}(q)-1\Big),
\end{equation}
where
$$\mathcal{I}(q)=\frac{2}{\pi}\int_{0}^{\pi/2}\frac{1}{1-f^{2}(q,\phi)}\left[\frac{\partial f(q,\phi)}{\partial\phi}\right]^{2}d\phi,$$
with $\mathcal{I}(q)\ge1$. Indeed, $\mathcal{I}(q)=1$ for $q=1$, so that the onion-like pattern recovers the ferromagnetic state along the $xy$-plane. In addition, $\mathcal{I}(q)>2$ whenever  $q>7$, so that such configurations (with $q>7$) cannot occur as ground-states, once their total energy is higher than the vortex one, as one can realize from Fig. \ref{Vortex-onion-Exch}.

\subsection{Magnetostatic term}
Once no effective magnetic charges appear in the vortex state, its total energy comes solely from exchange cost, computed above. The remaining patterns have magnetic charges lying in the torus surface and/or volume, making their energetics rather involved. Below, we shall present only the results and their main features, leaving the details for the respective Appendices. In every case, we start by determining the magnetostatic potential, from what the magnetic field emerges by $H=-\nabla\Phi_m$, as usual. Our procedure is based upon the evaluation of the 3D Green-function in terms of toroidal harmonics, involving Legendre functions of half-integer orders, $P^k_{n\pm 1/2}$ and $Q^k_{n\pm 1/2}$, of first and second kinds, respectively (we refer reader to Appendix {\bf B} for details).\\

Let us start by the simplest of such cases, the ferromagnetic (or single-domain, SD) along $z$-axis, $\vec{m}_{{\cal F}_z}=\hat{z}$, in which only $\sigma_m$ appears. We formally obtain (see Appendix {\bf C.1} for details):
\begin{equation}\label{EmagSDz}
E_{\text{mag}}^{\mathcal{F}_{z}}=\frac{128\mu_{0}M_{S}^{2}b^{3}}{9}\sinh^{2}\alpha_{0}\sum_{n=1}^{\infty}n^{2}P_{n-1/2}Q_{n-1/2}\left(Q^{1}_{n-1/2}\right)^{2},
\end{equation}
where we have omitted the superscripts $k=0$ in the Legendre functions and also defined $P_{\nu}^{\mu}(\cosh\alpha_{0})=P_{\nu}^{\mu}$ and $Q_{\nu}^{\mu}(\cosh\alpha_{0})=Q_{\nu}^{\mu}$, where $\cosh\alpha_{0}=R/r$ is the torus aspect ratio. Although we have explicitly assumed a bulk torus, the result above readily gives its hollow counterpart (we shall return to this point later).\\

Below, we present the magnetostatic contribution associated to onion-like patterns, whose special case $q=1$ represents the single-domain along the plane, SD$_{xy}$. Now, both surface and volumetric charges take place, making the determination of $\Phi_m$ very intricate. Even for this involved case, we have obtained a formal analytical expression for every $q$, as below:
\begin{eqnarray}\label{EmagOnion}
 & E_{\text{mag}}^{\mathcal{O}}=&\frac{16\mu_{0}M_{S}^{2}b^{3}}{9\pi^{2}}\sum_{k=0}^{\infty}\sum_{n=0}^{\infty}(-1)^{k}\epsilon_{n}\epsilon_{k}\frac{\Gamma\left(n-k+\frac{1}{2}\right)}{\Gamma\left(n+k+\frac{1}{2}\right)}P_{n-1/2}^{k}\times\nonumber\\
& & \times\left[Q_{n-1/2}^{k}\mathcal{A}_{n}\left(\mathcal{A}_{n}\mathcal{J}_{k}'-\mathcal{B}_{n}^{k}\mathcal{J}_{k}''\right)-\mathcal{B}_{n}^{k'}\left(\mathcal{A}_{n}\mathcal{J}_{k}''-\mathcal{B}_{n}^{k}\mathcal{G}_{n}^{k'}\right)\right],
\end{eqnarray}
with the definitions:
$$\mathcal{J}_{k}'=\int_{0}^{2\pi}\mathcal{J}_{k}(\varphi')m_{r}(\varphi') d\varphi',\qquad \mathcal{J}_{k}''=\int_{0}^{2\pi}\mathcal{J}_{k}(\varphi')\frac{\partial [m_{\varphi}(\varphi')]}{\partial\varphi'} d\varphi'$$
$$\mathcal{B}_{n}^{k'}=-\frac{3}{2}\int_{\alpha_{0}}^{\infty}\frac{Q_{n-1/2}^{1}(\cosh\alpha)}{\sinh\alpha}d\alpha, \qquad \mathcal{G}_{n}^{k'}=\int_{0}^{2\pi}\mathcal{G}_{n}^{k}(\varphi')\frac{\partial [m_{\varphi}(\varphi')]}{\partial\varphi'} d\varphi'.$$
Above, we have used the coordinates $(\alpha,\beta,\varphi)$, presented in Appendix {\bf A} (namely, we have that $R=b\tanh\alpha$, $r=b\sinh\alpha$, and $\phi=\varphi$), whereas ${\cal A}_n$, ${\cal J}_k(\varphi')$, ${\cal B}^k_n$ and, ${\cal G}^k_n (\varphi')$ are given in Appendix {\bf C.2}. Further evaluation of energy (\ref{EmagOnion}) demands numerical integration with $(R,r)$ as inputs.\\

By taking $q=1$ in Eq. (\ref{EmagOnion}) we obtain the magnetostatic energy associated to SD$_{xy}$, as below:
\begin{equation}\label{EmagSDxy}
E_{\text{mag}}^{\mathcal{F}_{xy}}=-\frac{32\mu_{0}M_{S}^{2}b^{3}}{9}\sum_{n=0}^{\infty}\epsilon_{n}\mathcal{A}_{n}^{2}P_{n-1/2}^{1}Q_{n-1/2}^{-1},
\end{equation}
where we have used the property $\frac{\Gamma\left(n-\frac{1}{2}\right)}{\Gamma\left(n+\frac{3}{2}\right)}Q^{1}_{n-1/2}=Q^{-1}_{n-1/2}$ (see Ref. \cite{Gradshteyn}, p. 959).\\

Let us return to the hollow torus, with thickness $t$. The potential $\Phi_m$ may be readily computed through linear superposition of those associated to two concentric tori with radii $r-t$ and $r$, having opposite magnetization on their surfaces. After that, an analogous procedure yields the magnetostatic energy. To make use of previous results, let us take coordinates $(\alpha,\beta, \varphi)$, so that two arbitrary tori, $A$ and $B$, are parametrized by $(\alpha_A, \beta, \varphi)$ and $(\alpha_B, \beta', \varphi')$, respectively. Then, we have $\cosh(\alpha_A)=R/r_A$ and $\cosh(\alpha_B)=R/r_B$, and by assuming $r_A>r_B$, we obtain, for the ferromagnetic state along $z$:
\begin{eqnarray} \label{EmagSDzhollow}
E_{\text{mag}}^{\mathcal{F}_{z}}\Big|_{\rm hollow}& = & \frac{128\mu_{0}M_{S}^{2}}{9}\sum_{n=1}^{\infty}n^{2}\left[b^{3}_{A}\sinh^{2}\alpha_{A}P_{n-1/2}^{A}Q_{n-1/2}^{A}\left(Q^{1A}_{n-1/2}\right)^{2}\right.\nonumber\\
& &  \left.-b^{3}_{B}\sinh^{2}\alpha_{B}P_{n-1/2}^{B}Q_{n-1/2}^{B}\left(Q^{1B}_{n-1/2}\right)^{2}\right]
\end{eqnarray}
whereas for that along the plane, we have:
\begin{eqnarray}\label{EmagSDxhollow}
E_{\text{mag}}^{\mathcal{F}_{xy}}\Big|_{\rm hollow} & = & -\frac{32\mu_{0}M_{S}^{2}}{9}\sum_{n=0}^{\infty}\epsilon_{n}\Biggl\{b^{3}_{A}P_{n-1/2}^{1A}Q_{n-1/2}^{-1A}\left[\cosh\alpha_{A}\left(\frac{Q_{n+1/2}^{2A}}{\sinh\alpha_{A}}-nQ_{n-1/2}^{1A}\right)-\frac{Q_{n-1/2}^{2A}}{\sinh\alpha_{A}}\right]^{2}\nonumber\\ & & -b^{3}_{B}P_{n-1/2}^{1B}Q_{n-1/2}^{-1B}\left[\cosh\alpha_{B}\left(\frac{Q_{n+1/2}^{2B}}{\sinh\alpha_{B}}-nQ_{n-1/2}^{1B}\right)-\frac{Q_{n-1/2}^{2B}}{\sinh\alpha_{B}}\right]^{2}\Biggl\}.
\end{eqnarray}
Above, we have defined $b_{A,B}=\sqrt{R^2-r^2_{A,B}}$, while $P^{\mu}_{\nu}(\cosh\alpha_{A(B)})=P^{\mu A(B)}_{\nu}$ and $Q^{\mu}_{\nu}(\cosh\alpha_{A(B)})=Q^{\mu A(B)}_{\nu}$. The second terms in these expressions, proportional to $b_B$, identically vanish as long as $r_B \to 0$ and $r_A\equiv r$, as expected. The same may be done for the onion-like state, but due to its length, we shall not quote it here.\\

\section{Numerical results and discussion}
Although we have obtained analytical expressions for the magnetostatic energy of the desired magnetization in the torus, Eqs. (\ref{EmagSDz})-(\ref{EmagSDxy}), an immediate analysis of their physical content is not clear due to their complicated forms. Therefore, to extract relevant results we should evaluate them numerically. For doing that, we employ the Fortran subroutine code DTORH1 \cite{DTORH}, which gives the values of $P^k_{n-1/2}(\cosh\alpha)$ and $Q^k_{n-1/2}(\cosh\alpha)$ for a given torus with radii $R$ and $r$. We also adopt the values of $A$ and $M_s$ for Permalloy, already given in Section 2.

\begin{figure}\begin{center}
 \includegraphics[scale=0.4]{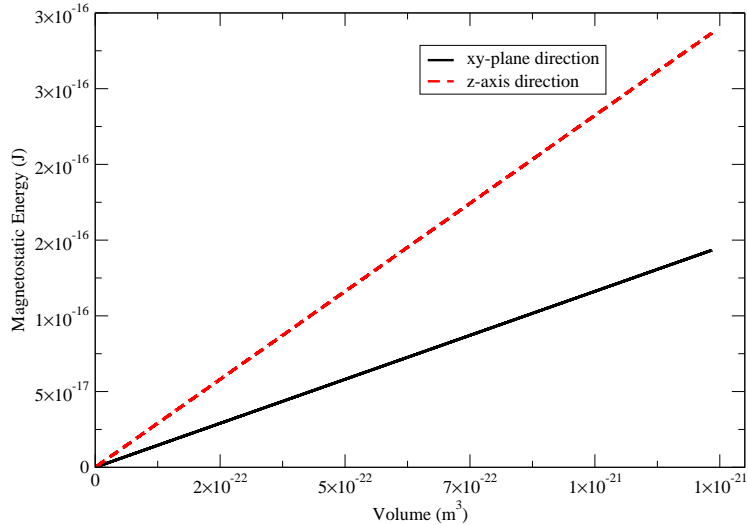}\hspace{1em}\caption{[Color online] The magnetostatic energy of the ferromagnetic states along $z$ and $xy$ against tori volume ($V=2\pi^{2}r^{2}R$). Above, we have fixed $r=5\,{\rm nm}$ whereas $R$ varies from $5.05$ nm to $2500$ nm. Clearly, SD$_{xy}$ is energetically favorable compared to its counterpart along $z$.}\label{figEmagferros}\end{center}
\end{figure}

\begin{figure}
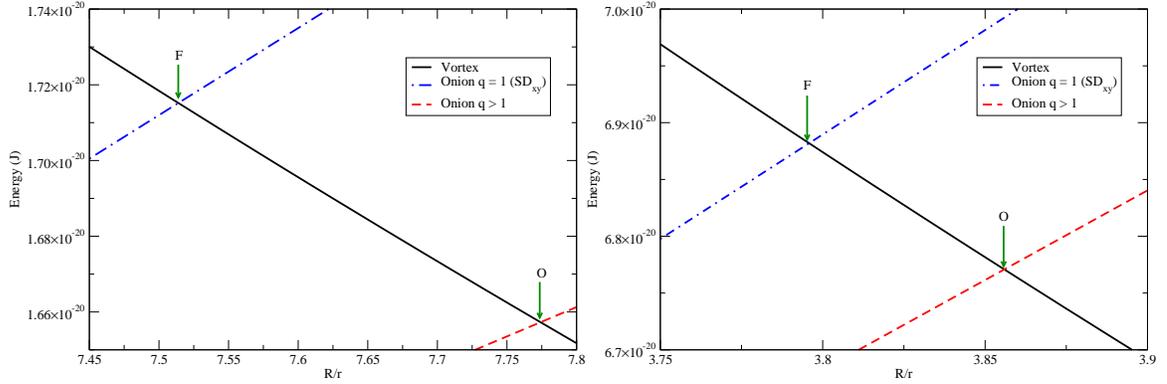
\begin{center}
 \includegraphics[scale=0.3]{qvraio1nm.eps}\includegraphics[scale=0.3]{qvraio2nm.eps}\caption{[Color online] The total energy corresponding to vortex, onion and SD$_{xy}$ ($q=1$) for $r=1\,{\rm nm}$ (left) and $r=2\,{\rm nm}$ (right). In each case, the point $O$ marks the critical $R/r$ value above which the onion state is no longer stable, say, its energy is higher than the vortex-state ($F$ is the point at which vortex and SD$_{xy}$ have the same net energy). The highest difference between the energies of the onion-like (with $q\approx1.7$) and SD$_{xy}$ states occurs for a torus with $r=1\,{\rm nm}$ and $R\approx8\,{\rm nm}$, and reads $\approx$ 3\% (see discussion in text).}\label{figEnmagonion}\end{center}
\end{figure}

\begin{figure}
 \begin{center}
  \includegraphics[scale=0.45]{SingDomainxy.eps}\includegraphics[scale=0.4]{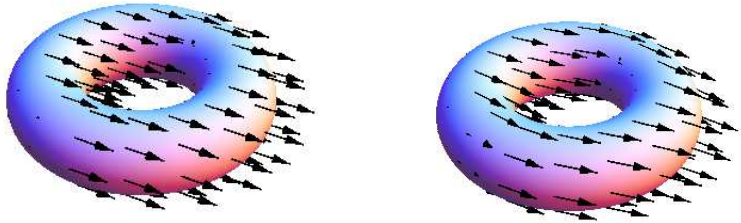}\caption{[Color online] Illustration of onion-like configurations with $q=1$ (ferromagnetic pattern; left) and $q=1.7$ (right). As also indicated by their respective energetics, these states are very similar each other (see text for further details).}\label{OnionDifq}
 \end{center}
\end{figure}

\begin{figure}\begin{center}
 \includegraphics[scale=0.4]{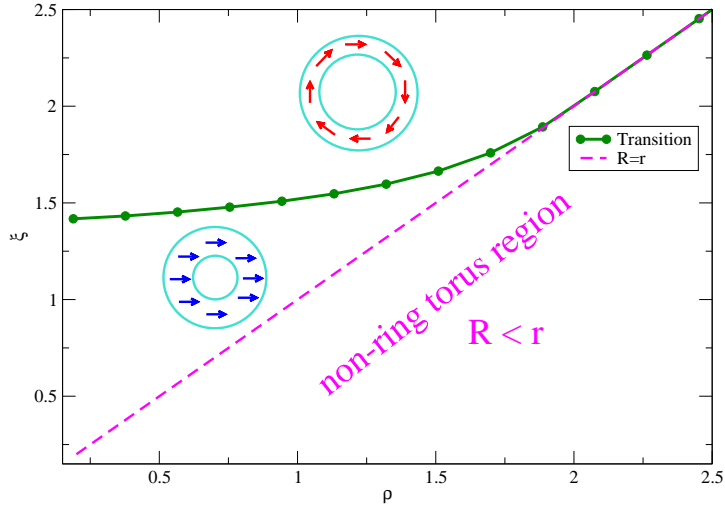}\caption{[Color online] State diagram for nanotori ($\xi=R/\ell_{ex}$ and $\rho=r/\ell_{ex}$). For every tori with $r\gtrsim10\,{\rm nm}$ ($\rho\gtrsim2$) the vortex pattern appears to be the ground-state. At smaler radii, $r\lesssim10\,{\rm nm}$, vortex remains the ground-state in tori with $R\gtrsim8\,{\rm nm}$ ($\xi\gtrsim1.5$), being substituted by SD$_{xy}$ at smaller $R$. The transition curve is precise within $\approx 3\%$ of confidence (for $\rho\lesssim1$ and $\xi\lesssim1.5$).}\label{statediagram}\end{center}
\end{figure}

First, we compare the energies associated to the ferromagnetic (single-domain) states along $z$ and $xy$, which may be summarized by the expressions below ($V=2\pi^{2}r^{2}R$ is the torus volume):
\begin{equation}
\frac{E_{\text{mag}}^{\mathcal{F}_{xy}}}{\mu_{0}M_{S}^{2}}\approx\frac{1}{8}V,\hspace{3em}\frac{E_{\text{mag}}^{\mathcal{F}_{z}}}{\mu_{0}M_{S}^{2}}\approx\frac{1}{4}V,
\end{equation}
clearly showing that SD$_{xy}$ costs nearly half of the energy from its counterpart along $z$ (Fig. \ref{figEmagferros}). Thus, hereon only this ferromagnetic pattern need to be considered. Next, recalling that it is a particular case of the onion-like state, with $q=1$, it is important to determine the values of this parameter that minimizes total energy for a given torus. This may be formally achieved by solving $\partial E/\partial q \equiv 0$, and the results are summarized in Fig. \ref{figEnmagonion}, where we present the net energy of vortex, onion and SD$_{xy}$ states against $R/r$. For each graphic, the onion state loses its stability at the point $O$, once it is more energetic than the vortex state for larger $R/r$. At this point, numerical results show that $q\approx 1.7$ minimizes the total energy in a torus with $R\approx8\,{\rm nm}$ and $r=1\,{\rm nm}$ (for larger $R$, the ground-state appears to be the vortex pattern). We can also note that the total energy of the onion state, at the point $O$, is only about 3.2\% less than its ferromagnetic (SD$_{xy}$) counterpart (indicated by the point $F$ in Fig. \ref{figEnmagonion}). This fact must be associated with the little deviation from parallel direction of the magnetic dipoles for $q=1.7$, as we can see in Fig. \ref{OnionDifq}. As $r$ is raised, such an energy difference between those configurations is diminished, as illustrated in Fig. \ref{figEnmagonion} (on the right) where $r=2\, {\rm nm}$ has been taken. Therefore, for saving time and computational efforts, we shall adopt $q=1$ as a very good approximation for evaluating the net energy, say, for our purposes it is suffice to compare vortex and SD$_{xy}$ states energetics.\\
To complete our analysis, we should compare vortex and SD$_{xy}$ energetics to decide what is the magnetization ground-state for a given torus. Figure \ref{statediagram} summarizes our main findings, where we have indicated the energetically favorable state for a given torus  aspect ratio, $R/r$. Considering only the physically suitable region, with $R>r$, we may realize that the vortex appears to be the ground-state in every torus with $r\gtrsim 10\,{\rm nm}$ ($\rho=r/\ell_{ex}\gtrsim 2$). In those tori with $r\lesssim 10\,{\rm nm}$ vortex still minimizes the total energy whenever $R\gtrsim 10\,{\rm nm}$ ($\xi=R/\ell_{ex}\gtrsim 2$) whereas SD$_{xy}$ is energetically favorable for smaller $R$. Mainly in the region $\rho\lesssim1$ and $\xi\lesssim1.5$ the true ground-state is not SD$_{xy}$ ($q=1$) exactly, but slightly different onion-like patterns with $1<q\lesssim1.7$, whose energies are, at best, 3.2\% below that the $q=1$ state. Therefore, the transition curve presents a confidence around this percentage in that region.\\

It is now instructive to compare our state diagram with its ring counterpart, for instance, that presented in Ref. \cite{Bellegia-etal-JMMM301-131-2006} (unfortunately, in this work authors do not provide the energy values for a better comparison among their and our results). Among other, a remarkable contrast concerns how small these geometries may support a vortex-like as the ground-state, which may occurs even in very small tori\footnote{We assume that our classical analysis remains approximately valid at such scales, of a few atoms along the internal diameter, where certainly quantum effects could be important. If we consider $r\gtrsim \ell_{\rm ex}\approx 5-6 \, {\rm nm}$, our statement about vortex stabilization is more reasonable within our approach.}, say, with $r\approx 1\, {\rm nm}$ and $R\gtrsim 8\, {\rm nm}$, whereas in a nanoring at least an external radius, $\gtrsim 10\,{\rm nm}$, with a central hole and thickness no less than $8\,{\rm nm}$ and $2\,{\rm nm}$ is demanded. More specifically, if $r=1\,{\rm nm}$ (thickness 2 nm), vortex pattern is stable for nanotori with volume $V\approx150\,{\rm nm}^{3}$, whereas for nanorings with the same thickness, this stability demands a volume not less than $V\approx225\,{\rm nm}^{3}$ (around 50\% higher than its nanotorus counterpart). Therefore, by virtue of its smooth curvature, the toroidal geometry offers a remarkable physical support to sustain miniaturization along with a single-vortex magnetization ground-state. In addition, thinking about an array of such nanotori, a system of practically non-interacting elements can be achieved provided that a separation $\gtrsim\ell_{ex}$ among neighbor nanotori is ensured. Although such a non-interacting feature may be obtained with nanorings, the nanotori has the advantage of minimizing considerably the magnetostatic fluctuations due to magnetic surface charges at the edges.\\

\begin{figure}
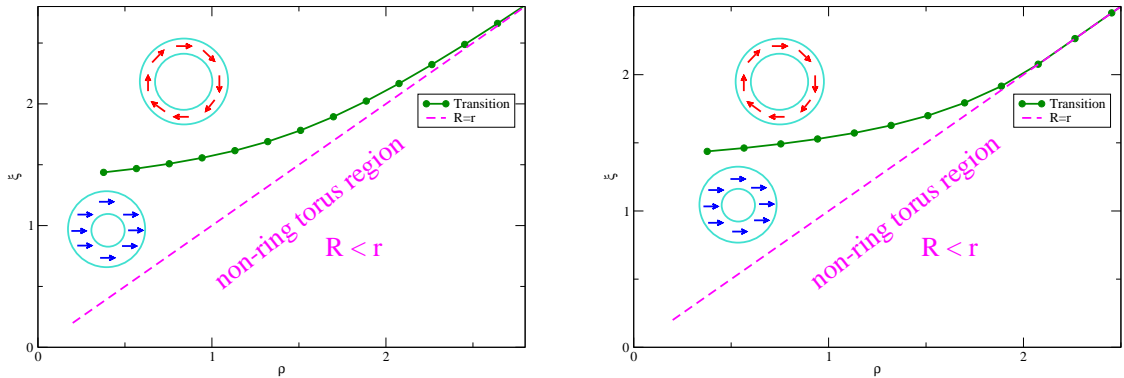

 \hskip -0.6cm \begin{center} \includegraphics[scale=0.3]{OnionDiagStFix.eps}\hskip 1cm
\includegraphics[scale=0.3]{OnionDiagStVar.eps}\caption{[Color online] State diagrams for the hollow torus. On the Left, a fixed thickness $t=1\,{\rm nm}$ and $r\in[2,20]\,{\rm nm}$ are taken; on the Right, the thickness varies as $t=r/2$. Namely, note that vortex stability tends to be (slightly) diminished as thickness is lowered, say, a thin hollow torus favors single-domain as its ground-state.}\label{DiagStHollow}
 \end{center}
\end{figure}

Now, we discuss upon the hollow torus case, whose vortex and single-domain energies are given by Eqs. (\ref{Eexvortex}) and (\ref{EmagSDxhollow}), respectively; two relevant state-diagrams are shown in Fig. \ref{DiagStHollow}. Namely, note that a higher value of $R$ is demanded for ensuring the vortex ground-state than in a bulk torus with the same $r$. Thus, in a hollow torus SD$_{xy}$ pattern appears to be more favorable as the torus thickness is diminished. However, at least for the cases considered, say, a torus with a fixed thickness $t=1\,{\rm nm}$ (Fig.\ref{DiagStHollow}, on the left), and with varying thickness according $t=r/2$ (Fig.\ref{DiagStHollow}, on the right), they are qualitatively similar to the bulk torus, with small quantitative differences.\\

\section{Conclusions and Prospects}
We have studied the energetics of a number of magnetization configurations lying in the geometry of a toroidal ferromagnet at nanoscales. From our analysis vortex-like emerges as the most prominent  of these patterns, being observed to remain stable even in very small tori, with $r\approx 1\,{\rm nm}$ and $R\gtrsim 8\, {\rm nm}$. Actually, onion-like configurations, with $1<q\lesssim1.7$, which are practically equivalent to SD$_{xy}$, takes place only for $\rho=r/\ell_{ex}\lesssim 1$ and $\xi=R/\ell_{ex}\lesssim 1.6$. This is an important result as long as miniaturization of basic magneto-electronic elements comprising a single-vortex concerns, once this configuration is stable in nanotori much smaller than in their ring and disc counterparts.\\

Under investigation are some possible mechanisms to induce switching between the two chirality states of such a vortex and, perhaps more interesting, the study of an array of such nanotori comprising a vortex in each one. In this situation, if the tori are spaced more than an exchange length from each other, we have a system of practically non-interacting elements. Indeed, even possible interactions induced by external agents (applied field, etc), which may deform the vortex, are expected to be minimized due to the toroidal smooth geometry. Additionally, such a smoothness is expected to prevent large magnetostatic fluctuations coming from surface magnetic charges, as may occurs in discs and rings straight corners. Furthermore, that weak self-interaction characteristic among the basic elements could make magnetic-vortex nanotori assembles as interesting systems in future targeted cancer-cell therapy \cite{Cancer-therapy}.\\

\centerline{\Large Acknowledgements}
\vskip .4cm
The authors are grateful to CAPES, CNPq and FAPEMIG (Brazilian agencies) for financial support.\\


{\centerline {\Large Appendices}}
\appendix

\section{Toroidal coordinate systems}
Besides the spherical-like coordinate system, presented in Section 2,  Eqs. (\ref{coord1}), the so-called toroidal coordinates are defined by means of the following parametric relations:
\begin{equation}\label{coord2}
x=b\frac{\sinh\alpha\cos\phi}{\cosh\alpha-\cos\beta}\,,\hspace{1.0em}
y=b\frac{\sinh\alpha\sin\phi}{\cosh\alpha-\cos\beta}\,,\hspace{1.0em}
z=b\frac{\sin\beta}{\cosh\alpha-\cos\beta}\,,
\end{equation}
where $b$ is a constant giving the radius of a circle at $z=0$ described by $\alpha\rightarrow\infty$, say, when $\alpha\rightarrow\infty$, we have $x=b\cos\phi$, $y=b\sin\phi$ and $z=0$; their ranges are $0\leqslant\alpha<\infty$, $0\leqslant\beta\leqslant2\pi$ and $0\leqslant\phi\leqslant2\pi$. On the torus surface we have:
\begin{equation}\label{ident}
b=\sqrt{(R+a)(R-a)},\hspace{1.0em}\cosh\alpha_{0}=\frac{R}{r},
\end{equation}
so that $b$ and $\alpha_{0}$ may be viewed as the \textit{geometric radius} and the \textit{eccentric angle}, respectively.\\

\section{Laplace Equation in Toroidal Coordinates; toroidal Green Functions}

The coordinate system (\ref{coord2}) is very useful because it has a known set of solution for the Laplace equation, where the method of separation of variables is valid (see Ref. \cite{Morse} p. 1302):
\begin{figure}\begin{center}
 \includegraphics[scale=0.6]{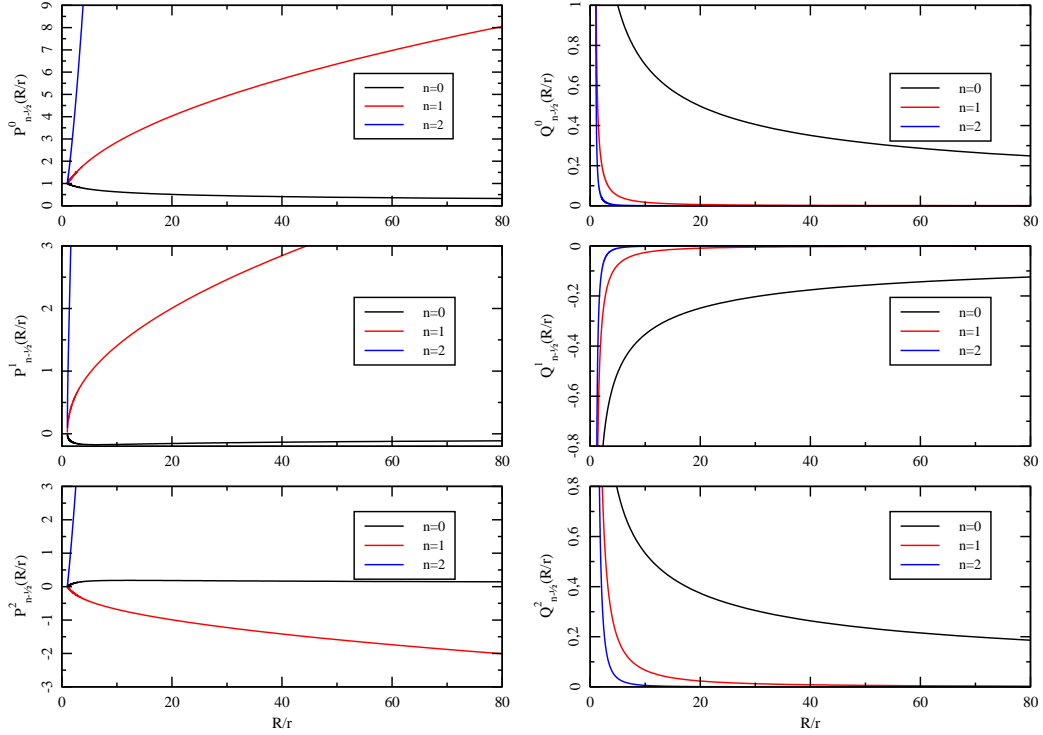}\caption{[Color online] The behavior of some toroidal functions against $R/r$.}\label{HarmTor}\end{center}
\end{figure}

 \begin{equation}
 \Phi(\alpha,\beta,\varphi)=\sqrt{\cosh\alpha-\cos\beta}A(\alpha)B(\beta)C(\varphi).
 \end{equation}
 where the functions $A$, $B$ and $C$ satisfy ($\zeta=\cosh\alpha$, and $n$, $k$ constants):
 \begin{equation}\label{dif1}
 (\zeta^{2}-1)\frac{d^{2}A}{d\alpha^{2}}+2\zeta\frac{dA}{d\alpha}
 -\left[\left(n^{2}-\frac{1}{4}\right)+\frac{k^{2}}{\zeta^{2}-1}\right]A=0,
 \end{equation}
 \begin{equation}\label{dif2}
 \frac{d^{2}B}{d\beta^{2}}+n^{2}B=0
 \end{equation}
 \begin{equation}\label{dif3}
 \frac{d^{2}C}{d\varphi^{2}}+k^{2}C=0.
 \end{equation}

 The functions $B(\beta)$ and $C(\varphi)$ are linear combinations of trigonometric functions, like below:

 \begin{equation}
 B_{n}(\beta)=f_{n}\cos n\beta+g_{n}\sin n\beta,
 \end{equation}
 \begin{equation}
 C_{k}(\varphi)=u_{k}\cos k\varphi+w_{k}\sin k\varphi,
 \end{equation}
 where $f_{n}$, $g_{n}$, $u_{k}$ and $w_{k}$ are constant coefficients to be determined from the boundary conditions. Since $\beta$ and $\varphi$ are variables ranging from 0 to $2\pi$,  the solutions must be periodic. This condition implies that $k=0,1,2,3,...$ and $n=0,1,2,3,...$, that is, invariance under rotations imposes $k$ and $n$ to be integers.

 The function $A(\alpha)$ are the Legendre functions of half-integer order of first kind $P^{k}_{n-1/2}$ and second kind $Q^{k}_{n-1/2}$, also called toroidal functions or toroidal harmonics, whose asymptotic behavior are given by (see Fig. \ref{HarmTor}):
 \begin{equation}\label{array}
 \begin{array}{rcl}
 \lim_{\alpha\rightarrow\infty}P^{k}_{n-1/2}(\cosh\alpha)=\infty\\
 \lim_{\alpha\rightarrow0}Q^{k}_{n-1/2}(\cosh\alpha)=\infty.
 \end{array}
 \end{equation}

The expansion $1/|\vec{r}-\vec{r}'|$ in terms of Green Function, according to the coordinate system (\ref{coord2}), is given by (See Ref. \cite{Morse}, p. 1304):

$$\frac{1}{|\vec{r}-\vec{r}'|}=\frac{\sqrt{(\cosh\alpha-\cos\beta)(\cosh\alpha'-\cosh\beta')}}{\pi b}\sum_{k=0}^{\infty}\sum_{n=0}^{\infty}(-1)^{k}\epsilon_{n}\epsilon_{k}\cos n(\beta-\beta')\cos k(\varphi-\varphi')$$
\begin{equation}\label{expand}
\times\frac{\Gamma\left(n-k+\frac{1}{2}\right)}{\Gamma\left(n+k+\frac{1}{2}\right)}P_{n-1/2}^{k}
(\cosh\alpha_{<})Q_{n-1/2}^{k}(\cosh\alpha_{>}),
\end{equation}
where $\epsilon_{n}=(2-\delta_{n,0})$, $\epsilon_{k}=(2-\delta_{m,0})$ and $\vec{r}^{'}$ is parameterized in terms of $(\alpha',\beta',\varphi')$. The fact of $P_{n-1/2}^{k}(\cosh\alpha)$ and $Q_{n-1/2}^{k}(\cosh\alpha)$ are related to $<$ and $>$, respectively, reflects the asymptotic behavior of these functions (see Fig. \ref{HarmTor}).\\

\section{Evaluation of the magnetostatic energy: expressions (\ref{EmagSDz}), (\ref{EmagOnion}) and (\ref{EmagSDxy})}

\subsection{Evaluation of Eq. (\ref{EmagSDz})} 
In the case of the SD$_z$, we have that $\vec{M}=M_{S}\hat{z}$, what gives:
\begin{equation}
\vec{M}\cdot\hat{n}=M_{S}\frac{\sinh\alpha\sin\beta}{\cosh\alpha-\cos\beta},
\end{equation}
so that only $k=0$ component contributes to the sum, and the integration in $\varphi$ is easily evaluated to give $2\pi$. Thus, the substitution of (\ref{expand}) in Eq. (\ref{potencialmag}) yields ($k=0$ superscripts has been omitted):

$$\Phi_{\mathcal{F}_{z}}=\frac{bM_{S}\sinh^{2}\alpha}{2\pi}\int_{0}^{2\pi}\frac{\sin\beta d\beta}{(\cosh\alpha-\cos\beta)^{5/2}}\sqrt{\cosh\alpha'-\cos\beta'}\sum_{n=0}^{\infty}\epsilon_{n}\cos n(\beta-\beta')$$
\begin{equation}
\times P_{n-1/2}(\cosh\alpha_{<})Q_{n-1/2}(\cosh\alpha_{>}).
\end{equation}

Now, using $\cos n(\beta-\beta')=\cos n\beta\cos n\beta'+\sin n\beta\sin n\beta'$, we obtain
$$\Phi_{\mathcal{F}_{z}}=\frac{bM_{S}\sinh^{2}\alpha}{2\pi}\sqrt{\cosh\alpha'-\cos\beta'}\sum_{n=0}^{\infty}\epsilon_{0}P_{n-1/2}(\cosh\alpha_{<})Q_{n-1/2}(\cosh\alpha_{>})$$
\begin{equation}\label{intermed}
\times \left[\cos n\beta'\int_{0}^{2\pi}\frac{\sin\beta \cos n\beta d\beta}{(\cosh\alpha-\cos\beta)^{5/2}}+\sin n\beta'\int_{0}^{2\pi}\frac{\sin\beta \sin n\beta d\beta}{(\cosh\alpha-\cos\beta)^{5/2}}\right]
\end{equation}

The first integral vanishes identically (integrand is odd under $\beta\rightarrow-\beta$), whereas the second one gives (Ref. \cite{Gradshteyn}, p. 961):
\begin{equation}\label{kind}
\int_{0}^{2\pi}\frac{\sin\beta\sin n\beta}
{(\cosh\alpha_{0}-\cos\beta)^{5/2}}d\beta=\frac{2n}{3}\int_{0}^{2\pi}\frac{\cos n\beta}{(\cosh\alpha_{0}-\cos\beta)^{3/2}}d\beta=-\frac{8n\sqrt{2}}{3\sinh\alpha_{0}}Q^{1}_{n-1/2}
\end{equation}
Now, once we are dealing with a toroidal shell, so that $\alpha'=\alpha=\alpha_{<}=\alpha_{>}\equiv\alpha_{0}$, we finally obtain:
$$\Phi_{\mathcal{F}_{z}}=-\frac{4\sqrt{2}bM_{S}\sinh\alpha_{0}}{3\pi}(\cosh\alpha_{0}-\cos\beta)^{1/2}\sum_{n=0}^{\infty}n\epsilon_{n}\sin n\beta$$
\begin{equation}\label{potz}
\times P_{n-1/2}(\cosh\alpha_{0})Q_{n-1/2}(\cosh\alpha_{0})Q_{n-1/2}^{1}(\cosh\alpha_{0}).
\end{equation}

Now, substitution of (\ref{potz}) in (\ref{Emag}) gives:
$$E_{\text{mag}}^{\mathcal{F}_{z}}=-\frac{4\sqrt{2}\mu_{0}M_{S}^{2}b^{3}}{3}\sinh^{3}\alpha_{0}\sum_{n=0}^{\infty}n\epsilon_{n}P_{n-1/2}Q_{n-1/2}$$
\begin{equation}
 \times Q^{1}_{n-1/2}\int_{0}^{2\pi} \frac{\sin\beta\sin n\beta d\beta}{(\cosh\alpha_{0}-\cos\beta)^{5/2}},
\end{equation}
which after integration by parts immediactly yields Eq. (\ref{EmagSDz}), as desired.

\subsection{Evaluation of Eqs. (\ref{EmagOnion}) and (\ref{EmagSDxy})}
For the onion-type patterns, we have:

\begin{equation}\label{escon}
 \vec{M}\cdot\hat{n}=M_{S}\frac{\cosh\alpha\cos\beta-1}{\cosh\alpha-\cos\beta}m_{r}(\varphi),
\end{equation}
and

\begin{equation}\label{divon}
 \vec{\nabla}\cdot\vec{M}=M_{S}\left(\frac{\cosh\alpha-\cos\beta}{b\sinh\alpha}\right)\frac{\partial m_{\varphi}(\varphi)}{\partial\varphi}.
\end{equation}
Now, the substitution of (\ref{escon})-(\ref{divon}) in (\ref{potencialmag}), we obtain:
$$\Phi_{\mathcal{O}}=\frac{bM_{S}}{4\pi^{2}}(\cosh\alpha'-\cos\beta')^{1/2}\sum_{k=0}^{\infty}\sum_{n=0}^{\infty}(-1)^{k}\epsilon_{k}\epsilon_{n}\frac{\Gamma\left(n-k+\frac{1}{2}\right)}{\Gamma\left(n+k+\frac{1}{2}\right)}P_{n-1/2}^{k}$$$$\times\Biggl\{\sinh\alpha_{0}Q_{n-1/2}^{k}\int_{0}^{2\pi}d\beta\frac{(\cosh\alpha_{0}\cos\beta-1)\cos n(\beta-\beta')}{(\cosh\alpha_{0}-\cos\beta)^{5/2}}\int_{0}^{2\pi}m_{r}(\varphi)\cos k(\varphi-\varphi')d\varphi$$
\begin{equation}
-\int_{\alpha_{0}}^{\infty}\frac{Q_{n-1/2}^{k}(\cosh\alpha)}{\sinh\alpha}d\alpha\int_{0}^{2\pi}\frac{\cos n(\beta-\beta')}{(\cosh\alpha-\cos\beta)^{3/2}}\int_{0}^{2\pi}\frac{\partial m_{\varphi}(\varphi)}{\partial\varphi}\cos k(\varphi-\varphi')d\varphi\Biggl\},
\end{equation}
with $P_{\nu}^{\mu}(\cosh\alpha_{0})=P_{\nu}^{\mu}$, and similarly for $Q$. Using the usual identity $\cos n(\beta-\beta')=\cos n\beta\cos n\beta'+\sin n\beta\sin n\beta'$, and integration by parts, we get:

$$\Phi_{\mathcal{O}}=\frac{2\sqrt{2}bM_{s}}{3\pi^{2}}\left(\cosh\alpha'-\cos\beta'\right)^{1/2}\sum_{k=0}^{\infty}\sum_{n=0}^{\infty}(-1)^{k}\epsilon_{n}\epsilon_{k}\cos n\beta'\frac{\Gamma\left(n-k+\frac{1}{2}\right)}{\Gamma\left(n+k+\frac{1}{2}\right)}P^{k}_{n-1/2}$$
\begin{equation}\label{poton}\times\left[Q_{n-1/2}^{k}\mathcal{A}_{n}\mathcal{J}_{k}(\varphi')-\mathcal{B}^{k}_{n}\mathcal{G}_{n}^{k}(\varphi'))\right],
\end{equation}
where
$$\mathcal{A}_{n}=\left[\cosh\alpha_{0}\left(\frac{Q_{n+1/2}^{2}}{\sinh\alpha_{0}}-nQ_{n-1/2}^{1}\right)-\frac{Q_{n-1/2}^{2}}{\sinh\alpha_{0}}\right],\hspace{4em}\mathcal{J}_{k}(\varphi')=\int_{0}^{2\pi}m_{r}(\varphi)\cos k(\varphi-\varphi')d\varphi$$
$$\mathcal{B}_{n}^{k}=-\frac{3}{2}\int_{\alpha_{0}}^{\infty}\frac{Q_{n-1/2}^{k}(\cosh\alpha)Q_{n-1/2}^{1}(\cosh\alpha)}{\sinh\alpha}d\alpha,\hspace{4em}\mathcal{G}_{n}^{k}(\varphi')=\int_{0}^{2\pi}\frac{\partial[m_{\varphi}(\varphi)]}{\partial\varphi}\cos k(\varphi-\varphi')d\varphi.$$

After a long and tedious algebraic manipulation, substitution of expression (\ref{poton}) in Eq.(\ref{Emag}) yields the desired Eq. (\ref{EmagOnion}), from what Eq. (\ref{EmagSDxy}) is readily obtained by setting $q=1$.

\end{document}